\documentclass[twocolumn]{revtex4-1}
\usepackage{amsmath,amsfonts,amssymb, amstext}
\usepackage{graphicx}
\begin{document}

\title{ On-chip all-silicon thermoelectric device}

\author{Antonella Masci}
\affiliation{Dipartimento di Ingegneria Informazione, Università di Pisa,
 Via G.Caruso, I-56122, Pisa-Italy}
 
\author{Elisabetta Dimaggio}
\affiliation{Dipartimento di Ingegneria Informazione, Università di Pisa,
 Via G.Caruso, I-56122, Pisa-Italy}
 
 \author{Giovanni Pennelli}
\affiliation{Dipartimento di Ingegneria Informazione, Università di Pisa,
 Via G.Caruso, I-56122, Pisa-Italy}


\maketitle

%\twocolumn[
 % \begin{@twocolumnfalse}
    
%\begin{abstract} 
The low thermal conductivity of silicon nanostructures, with respect to bulk silicon, opens excellent possibilities 
for thermoelectric applications because it will enable the use of silicon for the high efficient direct conversion 
of wasted heat  into electrical power. This paves the way for the application of silicon devices for energy
scavenging and green energy harvesting.
We present a device with a large number of nanostructures suspended on a silicon substrate. 
Top-down techniques for device fabrication based on advanced lithography and anisotropic etching will be discussed. 
FEM simulations were also carried out to analyze the temperature trend through the nanostructures.
%\end{abstract}

\section{Introduction}
\begin{figure}  [ht!]
\begin{center}
\includegraphics[scale=0.45]{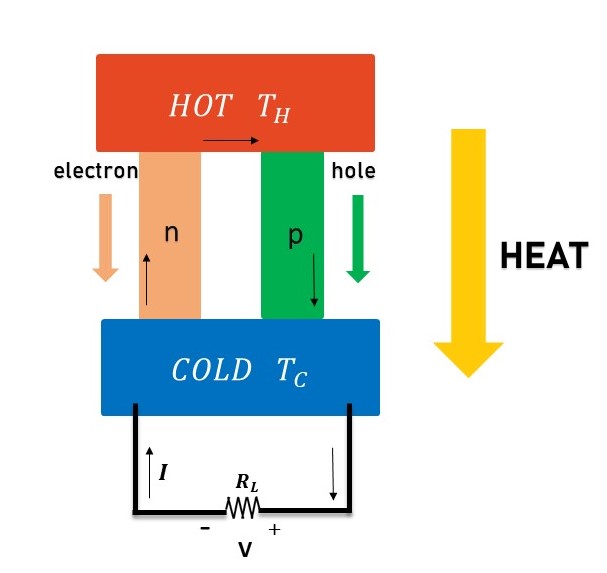}
\end{center}
\caption{ Sketch of a base cell of a thermoelectric generator with load $R_{L}$.} 
\label{fig:basecell}
\end{figure}

Heat is one of the most important energy sources on Earth, it represents
an inexhaustible and ecological form of energy. In a global context, where
the demand for electricity is always increasing, sources of sustainable energy
have gained more and more interest. 
Given its enormous potential, there is a strong interest in researching economic technologies for the direct 
conversion of wasted heat into electrical energy, which so far has found application only in a few sectors 
for the limitations imposed by currently available thermoelectric materials.  
The thermoelectric effect is exploited to convert the heat provided from a
source, which would otherwise be dissipated, into electricity useful for continuously powering small devices. The advantages of the 
thermoelectric generators (TEG) are comparable to those obtained from other renewable sources, but with the founding ideas of the circular 
economy: exploiting waste, or
already existing sources such as the sun, wind, water or heat to obtain electricity. The purposes of TEG devices are precisely to convert heat (poor energy) into electricity (precious energy) with high efficiency. 
The thermoelectric conversion is based on the Seebeck effect according to which, in a circuit consisting of semiconductors, a difference of temperature generates electricity. 
The unit cell of a thermoelectric generator\cite{mio-beilstein} is composed by two legs of materials with different Seebeck coefficient, placed between a hot and a cold source. These legs are electrically arranged in series and thermally in parallel. The heat flows through the two legs and determines a flow of charge carriers that generates an electrical power on the load $R_{L}$(Figure \ref{fig:basecell}).

%In a previous work, we demonstrated a strong reduction
%of the thermal conductivity in silicon nanomembranes 1 $\mu$m wide and 100 nm thick.
%In this work, we use the same nanostructures but the large 

\begin{figure*}  [ht!]
\begin{center}
\includegraphics[scale=0.45]{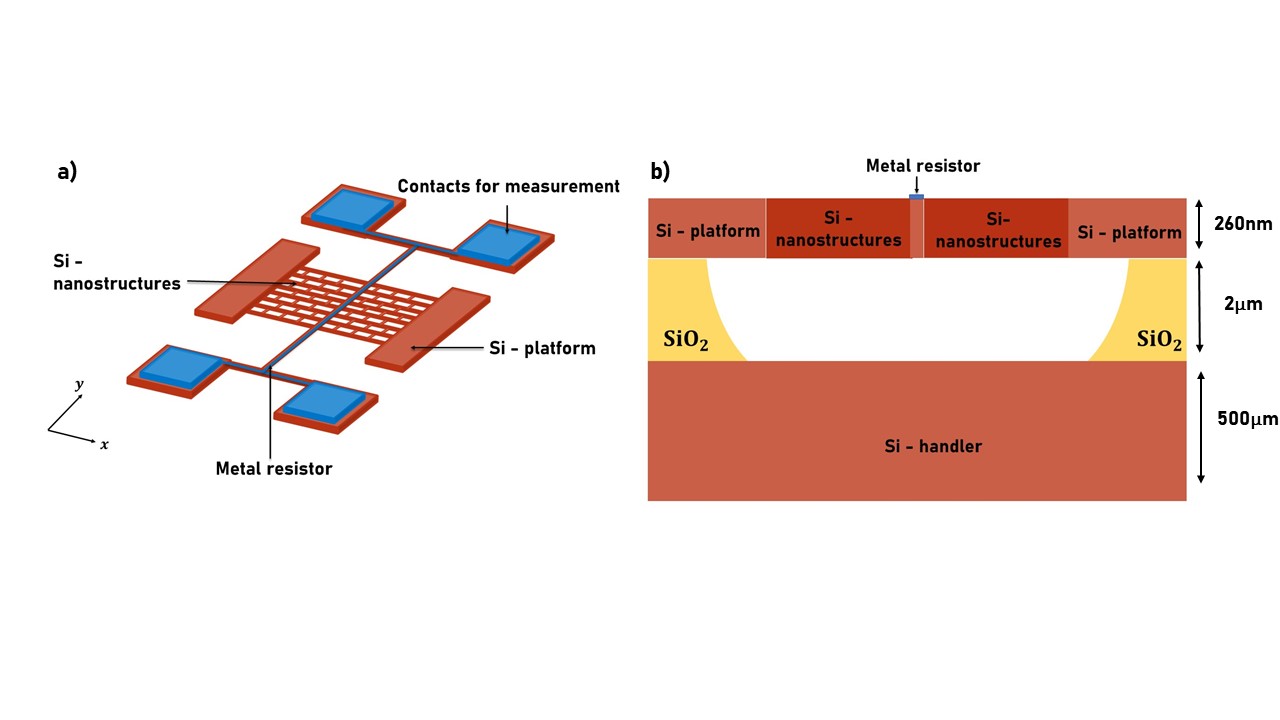}
\end{center}
\caption{ Sketches of the device. Panel a): an overall vision of the device. Panel b): a cross section in $x$ direction, showing
the suspended nanostructures.} 
\label{fig:sketch3D}
\end{figure*}

\begin{figure*}  [ht!]
\begin{center}
\includegraphics[scale=0.7]{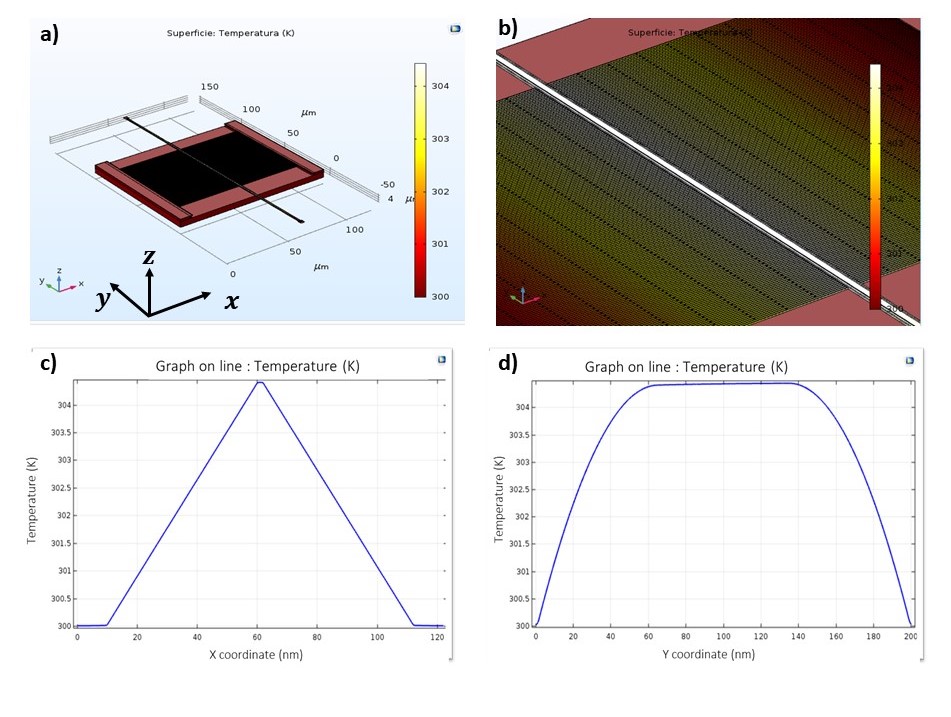}
\end{center}
\caption{FEM simulation results. Panel a): FEM simulation of temperature distribution for the device sketched in the figure\ref{fig:sketch3D}. Panel b): FEM simulation of temperature distribution (enlargement of the device).
Panel c): temperature trend along the nanostructures ( $x$ axis). Panel d): temperature trend along the central
silicon ribbon ($y$ axis), where there is the metal track acting as heater generator.} 
\label{fig:simul}
\end{figure*}
The efficiency of TEG devices is related\cite{heat-2008, hochbaum-2008} to the figure of merit, Z = $\dfrac{S^{2} \sigma}{k_{t}}$ : maximizing Z is equivalent to choose a material with a high electrical conductivity $\sigma$, a high Seebeck coefficient $S$ and a thermal conductivity $k_{t}$ as small as possible.
In this context, silicon, a bio-sustainable material, abundant on Earth, offers interesting properties\cite{mio-nanofili-doppi-2018}. Its chemical, physical and structural characteristics are well known due to its large diffusion in the nanotechnology and nano-electronics industry. However, bulk silicon,
compared to other materials used for thermoelectric applications, has a high thermal conductivity, equal to 148 W/(m K). The thermal conductivity 
is given by the sum of two contributions\cite{dresselhaus1,dresselhaus2} one due to charge carriers and the other to phonons. 
In bulk silicon, the greatest contribution is due to the phonons, therefore the thermal conductivity can be considered approximately equal to the phonon thermal conductivity.
Several works, present in the literature\cite{li-2003,li-lee-2012,mio-nanotechnologyII}, have 
shown that, by reducing down to the nanoscale the dimensions of silicon structures, the thermal conductivity results in a few W/(m K)\cite{mio-p+leg}.
In nanostructures, the dimensions of the devices become comparable with the phonon mean free path. This leads to a reduced phonon propagation, and 
therefore the thermal conductivity drops to very low values.
Our purpose is the development of a fabricating process for a thermoelectric device based on a large number of nanostructures\cite{mio-generatore-orizzontale}. A prototype of these thermoelectric generators was defined on top of a silicon on insulator wafer, together with an integrated metallic resistor used for the application of the three $\omega$ technique for the measurement of the thermal conductivity.
\begin{figure*} [ht!]
\begin{center}
\includegraphics[scale=0.45]{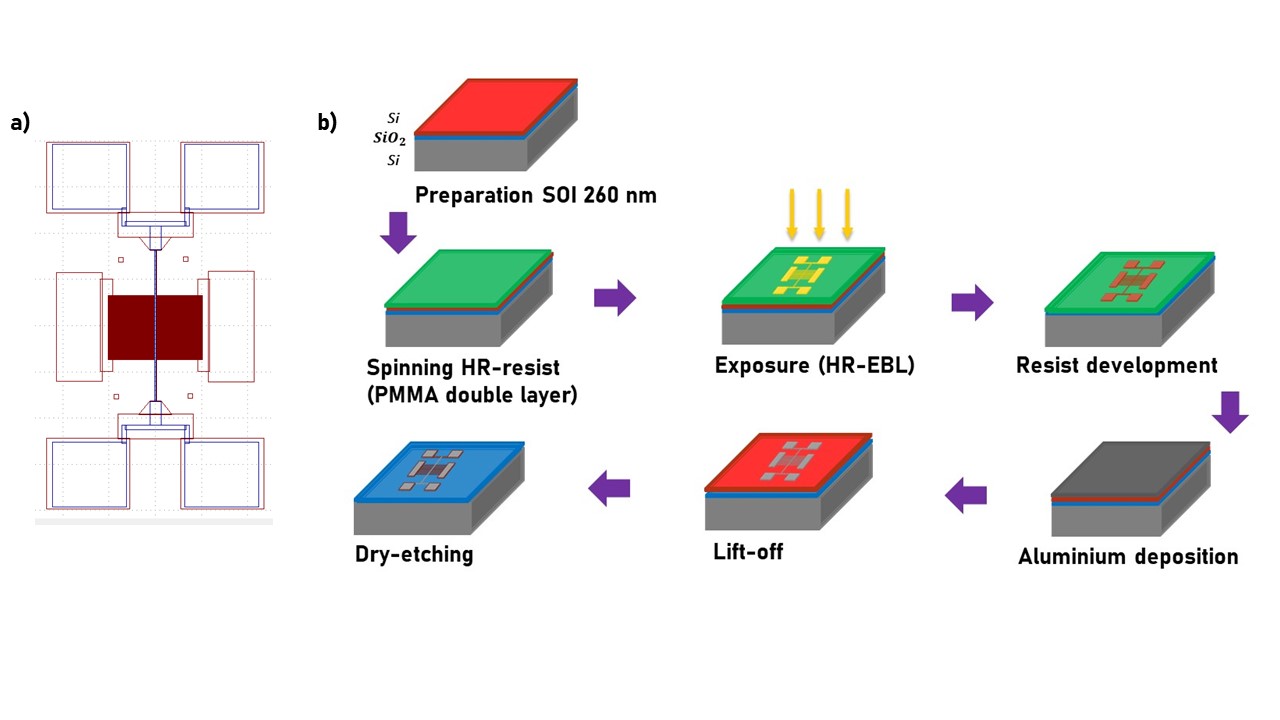}
\end{center}
\caption{Panel a): mask design of the device: red geometries represent the layout of the nanostructures, blue geometries represent
the metal track and the pads for the external connections. Panel b): sketches of the process steps for the fabrication of the
device.} 
\label{fig:process}
\end{figure*}

\section{Design of the device}
The device is based on a large array of nanostructures, fabricated on a Silicon-On-Insulator (SOI) wafer
with a top silicon layer 260 nm thick, a buried oxide layer 2 $\mu$m thick and an handler layer 
500 $\mu$m thick. The large array of nanostructures is defined on the top silicon layer by 
an high resolution lithographic step, followed by an anisotropic plasma etching step.
The nanostructures are silicon nanoribbons, narrow down to 100nm and as tall as the thickness
of the top silicon layer (260 nm). As the large side is perpendicular to the surface, this 
design allows the packing of a large number of nanostructures, limited by the width and the pitch
between the nanoribbons. 
As shown in Figure \ref{fig:sketch3D}, these nanostructures are connected to
each other to increase their mechanical stability, so that they result arranged in two large grids.
These two grids  are suspended between two lateral large silicon platforms. The two grids are
connected in the middle by a silicon ribbon, 2.4 $\mu$m wide a 200 $\mu$m long.
In the top of the central silicon ribbon, a metal track (metal resistor) is fabricated,
to be used for the thermal characterization of these grids. This central silicon ribbon is connected
to top and a bottom platforms, where metal contacts for external connections are fabricated.
Practically, on the central silicon ribbon a metal track (metal resistor), connecting the top and bottom
metal pads, has been fabricated.
The Joule heating of this metal resistor has been used for the generation of a heat flux
through the suspended nnanostructures. The generated heat is then dissipated in the substrate
through the two lateral silicon platforms.
The main goal of this work is to characterize the heat flux through the nanostructures,
in order to determine their thermal conductivity. However, the design and the main concept 
of this device can be easily applied to more complete on-chip devices for energy scavenging through 
direct thermal to electrical energy conversion.

\begin{figure*} [ht!]
\begin{center}
\includegraphics[scale=0.6]{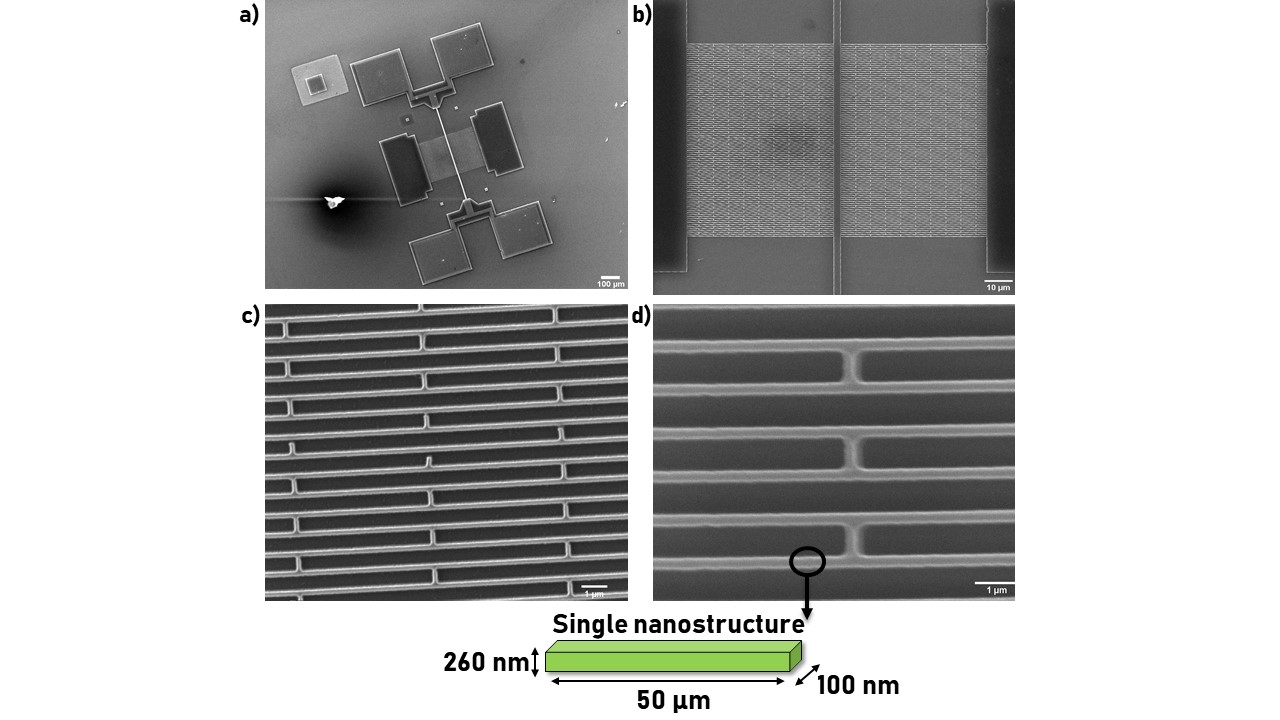}
\end{center}
\caption{SEM images of the device fabricated. In the bottom inset: a sketch of the nanostructure with its dimension.} 
\label{fig:SEM}
\end{figure*}
\subsection{Multiphysics simulation}
Finite Element (FEM) simulations have been performed for a correct design of the device. The 
software \textbf{COMSOL multiphysics} has been used for the simulations.
As known, this software allows to solve  the partial differential equations of thermal and electrical transport
taking into account complex geometries, as that of the proposed devices. A suitable meshing has been performed 
in order to have the right rounding precision in the simulations.
In order to achieve numerical values as close as possible to those of the experimental conditions, a multiphysic study 
has been performed: electrical transport has been considered in the metal track, where heat is generated by Joule
effect; the heat transport has been simulated from the central silicon ribbon, through the suspended nanostructures,
to the substrate where it is dissipated.
Figure~\ref{fig:simul} shows a typical simulation. Neumann boundary conditions have been fixed for the current
flux through the metal resistor; a Dirichlet boundary condition has been used to fix the temperature of the
bottom of the substrate. For the silicon pads and platforms a bulk thermal conductivity
of 148 W/(m K) has been considered; the thermal conductivity of the nanostructures has been fixed to 20 W/(m K),
similar to that already measured on suspended silicon nanomembranes with the large side parallel to the
substrate\cite{pennelli-dimaggio-nanomembrane-2018}.
The sketch of the simulated structure is shown in panel a) of Fig.~\ref{fig:simul}, and a typical result is
shown in panel b) where the temperature is reported with a color scale. Panel c) shows a temperature profile
along the nanostructures, taken between the suspending silicon platforms. Panel d) reports the temperature
profile  along the central silicon ribbon, from the top to the bottom contacts which have been fixed at
300 K.  This last graph shows that the temperature is almost constant in the middle of the nanostructure grids.
%Different values of thermal conducibility of silicon were used, in the silicon bulk equal to 148 W/(m K), 
%for the silicon of nanostructures equal to 20 W/(m K).
%For this work was performed multiphysics study:  to couple the electrical and the thermal physics with
%to purpose of evaluating the thermoelectric performances of the simulated
%device. The simulation allowed to get as close as possible to the experimental conditions, applying a current along the central metal and 
%xploiting the heating due to the Joule effect.
%The results led to a temperature drop confined to the extremities in the nanostructures. The figure \ref{fig:simul} shows the results of 
%the siulation and the temperature trend as a function of the x coordinate, parallel to the nanostructures and the coordinate y, 
%perpendicular to the nanostructures.

\section{Fabrication of the device}
The fabrication process is based on a top-down approach on a SOI (Silicon On Insulator) wafer, following
a planar strategy: the nanostructures are fabricated parallel to the surface of the silicon wafer. 
The process is sketched in Figure~\ref{fig:process}.
At first, an aluminum mask has been defined on the top silicon layer through high resolution electron beam lithography
(the design of the mask is shown in the left panel of Fig~\ref{fig:process}). To this end, standard PMMA
resist has been spun on the wafer; after e-beam exposure and development, an aluminum layer 50 nm thick has
been deposited by thermal evaporation, followed by lift-off in hot acetone.
The shaped aluminum has been used as a mask for the highly selective plasma etching in CF$_4$ atmosphere.
Hence, the width of the nanostructures (100 nm) is defined in the lithographic step, the height is that of the
top silicon layer (260 nm). After the plasma etching, the two grids of nanostructures result defined on the top silicon
layer, together with the silicon platform, the silicon ribbon in the  middle and the four areas on the top
and on the bottom, where metal pads have been fabricated. A second lithographic step, followed by gold
thermal evaporation (80 nm thick) and lift-off, has been used for the definition of the metal track (metal
resistor), exactly aligned on the silicon ribbon in the middle of the grids. In the same step, metal pads, 
connected with the metal resistor, have been defined. The four metal pads will allow external connections
for a four contact measurement of the resistance.
At the end, the grids of the nanostructures have been suspended by etching the buried SiO$_2$ layer 
through BHF (Buffered HF). The etching time has been calibrated for a complete removal of the oxide under the two silicon 
grids and the silicon ribbon in the middle, which therefore remained suspended. The
silicon platforms, which are very large with respect to the grids and to the central ribbon, remained
anchored to the substrate through the SiO$_2$ layer.
Several prototypes have been fabricated, to demonstrate the feasibility of the process.
Figure \ref{fig:SEM} shows some SEM images of one typical device. The images have been taken at different 
magnifications, to show the overall device (panel a)), the silicon grids (panel b) and c), with different magnification) 
and a detail of the nanostructures (panel c)). A sketch of a single nanostructure is also shown. 

\section{Discussion and future work}
Our work demonstrates the feasibility of a process for the on-chip fabrication of large arrays of silicon
nanostructures, which are interconnected and suspended on the silicon substrate. The grids are very
large, if compared with the dimensions of the nanostructures, hence can drive high electrical current/high
power to supply small sensor nodes. The device has been
designed to allow the propagation of the heat flux from the middle of the nanostructures to the lateral
supporting silicon platforms. Hence, a temperature difference can be supported, as demonstrated with
numerical (FEM) simulations. Therefore, this structure is suitable for thermoelectric generator devices, 
which can exploit a heat source to be applied to the central ribbon.
In the specific case of this work, the fabricated devices can be used for the measurement of the thermal 
conductivity of the grids of nanostructures. Future work will consist in applying an all-electrical technique, such as the 
three $\omega$ method\cite{mio-misura-treomega}, for thermal conductivity measurement: to this end, it will be exploited
the Joule heating achieved by the bias of the metal resistor fabricated at the center of the grids.
Moreover, the fabrication process will be improved to achieve nanostructures 100 nm wide
and 1 $\mu$m tall. To this end, a deep reactive plasma etching will be implemented.

\addcontentsline{toc}{chapter}{References}
%\nocite{*}
%\bibliographystyle{unsrt}
\bibliography{Si-thermoelectric_masci-dimaggio-pennelli.bib}

\end{document}